\documentclass[a4paper,11pt]{article}
\pdfoutput=1
\usepackage{jheppub}

\usepackage{amsmath,amssymb,amsfonts,cases}
\usepackage{amsthm}
\usepackage{graphicx,caption,subcaption}
\usepackage{dsfont}
\usepackage{slashed}
\usepackage{comment}
\usepackage{soul}
\usepackage{float}
\usepackage{wrapfig}
\usepackage{mathtools}
\usepackage{tikz}
\usepackage[compat=1.1.0]{tikz-feynman}
\usepackage{slashed}

\usepackage{tikz-cd}
\usepackage{physics}
\usepackage{simpler-wick}
\usepackage[inline]{enumitem}

\def\be{\begin{equation}}
\def\ee{\end{equation}}

\def\bal{\begin{equation}\begin{aligned}}
\def\eal{\end{aligned}\end{equation}}

\def\d{{\rm d}}

\def\myconst{{\kappa_d}}

\def\comment#1{}

\renewcommand{\pb}{\mathbf{p}} 
\newcommand{\half}{\frac{1}{2}}

\title{\LARGE A Nonlocal Schwinger Model}
\author{Ludovic Fraser-Taliente$^\dagger$,}
\author{Christopher P.\ Herzog$^*$,}
\author{and Abhay Shrestha$^*$}

\affiliation{
${}^*$Department of Mathematics, King's College London,   Strand, London, WC2R 2LS, UK \\
${}^\dagger$Physics Department, Oxford University, 
Oxford, OX1 3PU, UK}

\abstract{ 
We solve a system of massless fermions constrained to two space-time dimensions interacting via a $d$ space-time dimensional Maxwell field.  Through dimensional reduction to the defect
and bosonization, the system maps to a massless scalar interacting with a nonlocal Maxwell field through a $F \phi$-coupling.  
The $d=2$ dimensional case is the usual Schwinger model where the photon gets a mass.  More generally, in $2<d<4$ dimensions,
the degrees of freedom map to a scalar which undergoes a renormalization group flow; in the ultraviolet, the scalar is free, while in the infrared it has scaling dimension $(4-d)/2$.  The infrared is similar to the Wilson-Fisher fixed point, and the physically relevant case $d=4$ becomes infrared trivial in the limit of infinite ultraviolet cut-off, consistent with earlier work on the triviality of conformal surface defects in Maxwell theory.
}

\makeatletter
\def\@fpheader{\vspace{0cm}}
\makeatother

\begin{document}
\maketitle


\section{Introduction}

In arguing that gauge invariance does not necessarily imply the existence of massless particles, Schwinger solved the example of 1+1 dimensional massless electrons propagating in an electromagnetic field \cite{Schwinger:1962tp}.  
These days, the model is often celebrated as a solvable example that shares many features with QCD.  
In the limit where the coupling to the electrons is removed, the linear dependence of the electromagnetic potential with distance means the theory is confining and Wilson loops obey an area law.  Restoring the coupling, heavy probe ``quarks'' are screened by the massless fermions and the area law turns into a perimeter law.

Although the original work proceeds differently, a modern bosonization perspective makes it straightforward to see that the photon picks up a mass proportional to the coupling constant \cite{ZinnJustin}.  
Here we present a family of solvable models that generalize Schwinger's original construction.  Instead of coupling our two dimensional massless electrons to a two dimensional photon, we imagine the electrons travel along a wire embedded in $d$-dimensional space-time with interactions mediated by $d$-dimensional photons.  In our formalism, we are able to let $d$ be any real number, not just a positive integer.  Curiously, $d=2$ remains the only case where the photon gets a mass.  The cases $d\geq 4$ and $d<2$ suffer various complications and pathologies, while in the range $2<d<4$, we find a renormalization group flow between two conformal field theory fixed points, from a standard free scalar in the UV to a generalized free scalar field in the IR.

The motivation for our work comes from defect conformal field theory studies and the claim that the set of conformal boundary conditions on defects in free conformal field theories is very limited.  If the bulk is a free scalar field, essentially all defects with codimension greater than one are ``trivial'' \cite{Lauria:2020emq}.  More recently two of us \cite{Herzog:2022jqv} showed the same for surface defects in four dimensional Maxwell theory.  Trivial here means that the bulk fields can couple only to generalized free fields on the defect.  

The question remains, however, what does happen when one couples charged degrees of freedom on a surface defect to four dimensional Maxwell theory.  Are generalized free fields involved?  Does a mass gap appear?
Our nonlocal Schwinger model exhibits renormalization group behavior reminiscent of  QED or $\phi^4$ field theory in the epsilon expansion. In $4-\epsilon$ dimensions, there is a  renormalization group flow between a free theory at short distance and theory with anomalous dimensions at long distance.  In precisely four dimensions, the two fixed points merge.  The theory requires the introduction of a short distance cut-off which immediately spoils conformal invariance and can introduce a Landau pole.  
In the infrared, the effective coupling flows to zero consistent with the ``triviality'' demonstrated in \cite{Herzog:2022jqv}.\footnote{%
As we will discuss more later, the behavior in our case is similar to that observed for a two dimensional non-linear sigma model on a surface defect coupled to a six dimensional free scalar theory \cite{Cuomo:2023qvp}. 
} 

Not surprisingly, closely related variants of our model have appeared in the condensed matter literature \cite{schulz1993wigner,wang2001coulomb,inoue2006conformal,naon2005conformal}, dubbed long-range Luttinger liquids. 
In these works, the electromagnetic field is traded for a long distance $1/r^\beta$ type interaction between the electron charge densities.  
 Indeed, the Coulomb potential for a point charge in $d-1$ spatial dimensions behaves as $1/r^{d-3}$, suggesting the relation $\beta=d-3$ between refs.\ \cite{naon2005conformal,inoue2006conformal} and our work, while refs.\ \cite{schulz1993wigner,wang2001coulomb} focus on the specific case $\beta = 1$.  
 The relationship however is not clear cut.  
These Luttinger type models explicitly break Lorentz invariance.  Although the charged degrees of freedom obey a linear dispersion relation, their speed is much slower than that of light. Further the model is on a lattice where the lattice spacing plays an important role, both as a UV regulator and also in giving a special role to the Fermi momentum $k_F$.
More recently, ref.\ \cite{Menezes:2016euw} considered precisely our case where $d=4$ and argues that the low energy description is a Luttinger liquid.  Although the analysis there is flawed,\footnote{%
 Personal communication N.~Menezes.} the conclusions echo the earlier work on long-range Luttinger liquids we just summarized.

Another closely related work is ref.\ \cite{Gorbar:2001qt} where the authors analyze the cases $d=3$ and $d=4$ using the Schwinger-Dyson equations in an improved ladder approximation.  Curiously, they find evidence that the fermion can get a mass although they rightly worry if an extension of the Coleman-Mermin-Wagner (CMW) theorem might forbid the breaking of chiral symmetry that leads to their mass.  
Such worries in the past have led to the lore that these types of ladder approximations are not well suited for fermions in one dimension \cite{giamarchi2003quantum,solyom1979fermi}.
Indeed, in recent work on CMW for defects \cite{Cuomo:2023qvp}, the case where only defect operators transform under the symmetry is a particular simple case where CMW should apply without modification.   

In our case with $N_f$ fermions and $U(N_f)_L \times U(N_f)_R$ chiral symmetry, we should divide the group into an overall $U(1)_L \times U(1)_R$, the axial part of which is anomalous and the vector part of which is gauged, 
and a residual $SU(N_f)_L \times SU(N_f)_R$. 
Consistent with the CMW theorem, in our exact solution, we find no such evidence for a fermion mass or spontaneous  breaking of the $SU(N_f)$ groups.
Instead our theory is infrared trivial for $d=4$ with infinite UV cutoff, similar to the behavior of $
\phi^4$ field theory or QED in four dimensions. 
In $d=3$, we find a generalized free scalar field in the IR, and no obvious evidence that the $SU(N_f)$ groups are broken.

Moving farther afield, other works have considered more general versions of mixed dimensional QED, where the electron is constrained to $p$ dimensions and the photon to $d>p$.  
 Kotikov and Teber  for instance have many works investigating mixed dimensional QED, also called RQED$_{d,p}$, from a perturbative perspective \cite{Kotikov:2013eha,Teber:2014hna,Teber:2012de}.
Heydeman et al.\  on the other hand have looked at these theories through the lens of conformal symmetry \cite{Heydeman:2020ijz} and confinement \cite{Heydeman:2022yni}.
Remarks in \cite{Heydeman:2020ijz} about marginality failing
for the $(p=2, d=4)$ case foreshadow the triviality result
 \cite{Herzog:2022jqv} mentioned above and underline the need to understand exactly what happens in this case.\footnote{%
Other related works are ref.\ \cite{Bellucci:2019viw} where fermions on a spatial topology  $\mathbb{R}^p \times (S^1)^q$ are considered and the thesis \cite{SanchezMonroy:2017bvu} which makes the observation of a continuous family of Schwinger-like models from continuation in $d$.  Refs.\ \cite{Narayanan:2010zg} and \cite{Narayanan:2009ag} consider the problem of 2d fermions in respectively
3d and 4d Yang-Mills theory.
}

 An outline of our work is as follows.
In section \ref{sec:setup}, based on Chapter 2 of \cite{Shrestha:2023hvk}, we set up the scalar and spinor versions of the problem and discuss how bosonization relates them.  
In the fermionic case, 
the reduction from $d$ dimensions to two dimensions is similar to the account in ref.\ \cite{Gorbar:2001qt} while the bosonization is a  generalization of the textbook discussion
\cite{ZinnJustin}.
The section concludes by demonstrating identical effective actions for both models, in each case a free massive scalar field with a modified kinetic term and no potential. Section \ref{sec:analysis} is then a detailed analysis of this kinetic term.  In $d=2$, we recover the familiar Schwinger model, while for $2<d<4$, we find a renormalization group flow from a free field in the UV to a generalized free field in the IR.  We then discuss in brief pathologies that occur in more general cases -- tachyons and negative spectral density.  The case $d=4$ requires a UV cutoff and special treatment.
Section \ref{sec:loopandcondensate} discusses Wilson and Polyakov loops
in our family of models.
Finally in section \ref{sec:discussion}, we conclude with a discussion.
Several appendices bury technical details.  Appendix \ref{app:conventions} gives our conventions for spinors.
 In appendix \ref{app:regulator}, we discuss an alternate UV regulator for the special case $d=4$, replacing a hard momentum cut-off with the same Gaussian regulator used by \cite{Gorbar:2001qt}, which eliminates the Landau pole.  
 Appendix \ref{app:sphereW} rederives the sphere free energy for generalized scalar fields.  In section \ref{sec:analysis}, we use these sphere free energies to argue that the renormalization group flow in $2<d<4$ has the expected monotonicity properties.

\section{Coupling Maxwell Theory to Charged Matter on a Surface}
\label{sec:setup}

We start with Maxwell theory on a $d$-dimensional manifold $M$ and a field theory with a conserved current on a two dimensional sub-manifold $N \subset M$.  
Although we say manifold, we will always take $M = {\mathbb R}^{1,d-1}$ and $N = {\mathbb R}^{1,1}$.  If a general point $x = ({\bf x}, y)$, then $N$ is located at $y=0$.  
We couple the two theories together, making use of the conserved current $J^a$ of the two dimensional theory. 
The classical action for such a construction looks as follows:
\be
\label{action}
S =  - \frac{1}{4}  \int_{M} \d^d x \, F_{\mu\nu} F^{\mu\nu} +   \int_{N} \d^2 {\bf x} \left( g J^a A_a+ {\mathcal L} \right) \ .
\ee
Here ${\mathcal L}$ is the Lagrangian density for the 2d field theory and $g \in {\mathbb R}$ is the coupling strength.  As usual $F_{\mu\nu} = \partial_\mu A_\nu - \partial_\nu A_\mu$.  
In what follows, we will look at two examples of ${\mathcal L}$, a massless free Dirac spinor and a massless free real scalar:
\be
{\mathcal L}_f =  i \bar \psi \slashed{\partial} \psi \  , \; \; \;
{\mathcal L}_s = -\frac{1}{2} (\partial^a \phi)(\partial_a \phi) \ .
\ee
(See appendix \ref{app:conventions} for spinorial and other conventions.)
The free fermion has the Noether current
$J^a = \bar \psi({\bf x}) \gamma^a \psi({\bf x})$, while the scalar has the topological current $J^a = \epsilon^{ab} \partial_b \phi({\bf x})$.  
As we will see later, the two examples are related by bosonization.

Before starting the analysis, it should be noted that the action (\ref{action}) is gauge invariant.  
In the case of the fermion, 
under gauge transformations given by $A_\mu \to A_\mu + \partial_\mu \chi$, $\psi \to e^{i g \chi} \psi$, and $\bar \psi \to \bar \psi e^{-i g \chi}$, the action is invariant.  In the case of the scalar, assuming $\phi \to \phi$ under the transformation, 
the change in the action can be written as a divergence on the defect through the use of off-shell current conservation $\partial_a J^a({\bf x}) = 0$, and thus gauge invariance is guaranteed through appropriate boundary conditions as $|{\bf x}| \to \infty$.

\subsection{Bosonization}

The two actions, the one with the free fermion and the one with the free scalar, are related by bosonization.
For the free theories, there is a relation at the level of the operators, which we can check using Wick's Theorem, starting with the
propagators of the fundamental fields.
The propagator for the scalar has the form
\be
\langle \phi({\bf x}) \phi({\bf 0}) \rangle = -\frac{1}{4\pi} \ln {\bf x}^2 \Lambda^2 \ , 
\ee
where $\Lambda$ is a UV regulator.  The propagator for the fermion on the other hand is
\be
\langle \psi({\bf x}) \bar \psi({\bf 0}) \rangle = - \frac{i}{2\pi} \frac{ \gamma_a x^a}{{\bf x}^2}  \ ,
\ee
which follows from acting on $\langle \phi({\bf x}) \phi({\bf 0}) \rangle$ with the Dirac operator $i \slashed{\partial}$.

Important for us is the relation between the currents.
If we consider the two-point functions of the currents, we find  (see for example \cite{ZinnJustin} or \cite{Hosotani:1998za}) 
\be
\langle J_s^a({\bf x}) J_s^b({\bf 0}) \rangle = \frac{1}{2\pi} I^{ab}({\bf x}) \ , \; \; \;
\langle J_f^a({\bf x}) J_f^b({\bf 0}) \rangle = \frac{1}{2 \pi^2} I^{ab}({\bf x}) \ ,
\ee
where
\be
I^{ab}({\bf x}) \equiv \frac{1}{ {\bf x}^2} \left(\eta^{ab} - 2 \frac{{\bf x}^a {\bf x}^b}{{\bf x}^2} \right) \ ,
\ee
which suggests the identification $J_s = \sqrt{\pi} J_f$.  Thus whatever physics we find in the scalar case, we should find an analog for the fermion with the rescaling of the coupling $g \to g/ \sqrt{\pi}$.

\subsection{Scalars}

We start by analyzing the simpler scalar case ${\mathcal L} = {\mathcal L}_s$, simpler because the action is quadratic in the fields.  
It is useful at this point to fix a gauge, to which end we modify the action by the gauge fixing term 
\be
S \to S -\frac{1}{2 \zeta} \int_M \d^d x (\partial_\mu A^\mu)^2 \ ,
\ee
where the gauge fixing parameter $\zeta$ is a real number.
After an integration by parts, the classical action becomes
\begin{eqnarray}
\label{scalarS}
S_s &=&  \int \d^d x \, \left(  - \frac{1}{4}  F_{\mu\nu} F^{\mu\nu}  - \frac{1}{2\zeta} (\partial^\mu A_\mu)^2 \right) \nonumber \\
&& \; \; \; \; + \int \d^2 {\bf x} \left(  \frac{g}{2} \epsilon^{ab} F_{ab}\phi  -\frac{1}{2} (\partial^a \phi)(\partial_a \phi) \right) \ .
\end{eqnarray}
Note we can also add an $\epsilon^{ab} F_{ab}$ term to the action, which has the effect of shifting $\phi$ by a constant and can thus be undone by a field redefinition.
To facilitate path integral manipulations, we write the interaction in a four dimensional form using a smearing function $f(y) = \delta(y)$, 
which for now is just the Dirac delta function.  In a major case of interest $d=4$, a UV divergence appears associated with the vanishing transverse size of the defect.  In this case, it will be useful to have a more general form for $f(y)$ to regulate the divergence.
We have then
\begin{eqnarray}
S_s &=&  \int \d^d x \, \left(  - \frac{1}{4}  F_{\mu\nu} F^{\mu\nu}  - \frac{1}{2\zeta} (\partial^\mu A_\mu)^2 + \frac{g}{2} f(y) \epsilon^{ab} F_{ab}(x) \phi({\bf x}) \right) \nonumber \\
&& \; \; \; \;  -\frac{1}{2}  \int \d^2 {\bf x}(\partial^a \phi)(\partial_a \phi) \ .
\end{eqnarray}

We find it simpler to work in a Fourier transformed language.  The action in Fourier space is
\begin{eqnarray}
S_s &=& \int \frac{\d^d p}{(2\pi)^d} \left[ -\frac{1}{2}\left(p^2 \eta_{\mu\nu} - (1-\zeta^{-1}) p_\mu p_\nu \right) \tilde A^\mu (p) \tilde A^\nu(-p)
+i g  \epsilon^{ab} p_a \tilde A_b (p) \tilde \Phi(-p) \right] \nonumber \\
&&  - \frac{1}{2}\int \frac{d^2 {\bf p}}{(2\pi)^2}  \, {\bf p}^2 \tilde \phi({\bf p}) \tilde \phi(-{\bf p}) \ , 
\end{eqnarray}
where $\Phi(x) \equiv \phi({\bf x}) f(y)$.  
We proceed by first integrating out the transverse gauge field $\tilde A^i$, treating the parallel field $\tilde A^a$ as a source.  The equation of motion for $A^i$ yields
\be
\tilde A^i(p) = p_a \tilde A^a(p) p^i (1- \zeta^{-1}) \frac{1}{p^2 - p^j p_j (1-\zeta^{-1})} \ , 
\ee
and the action reduces first to
\begin{eqnarray}
S_s &=& \int \frac{\d^d p}{(2\pi)^d} \left[ - \frac{1}{2} \left( p^2 \eta_{ab} - \frac{p_a p_b (1-\zeta^{-1})}{1- \frac{p^k p_k}{p^2} (1-\zeta^{-1})} \right) \tilde A^a(p) \tilde A^b(-p) + i g \epsilon^{ab} p_a \tilde A_b(p) \tilde \Phi(-p) \right] \nonumber \\
&& - \frac{1}{2} \int \frac{\d^2 {\bf p}}{(2\pi)^2} {\bf p}^2 \tilde \phi({\bf p}) \tilde \phi(-{\bf p}) \ .
\end{eqnarray}
We next integrate out the parallel components of the photon $\tilde A^a$.  The equation of motion yields
\be
\tilde A^a(p) = -i \frac{g }{p^2}p_b \epsilon^{ba} \tilde \Phi(p) \ .
\ee
Note that $p_a \tilde A^a(p) = 0$.  
The action further reduces to\footnote{To make a connection with how we deal with the spinor case later on, note this result could also be produced by replacing the $d$-dimensional Maxwell field with a  nonlocal Maxwell field $B_a$ in two dimensions, and then integrating this nonlocal Maxwell field out.
}
 \begin{eqnarray}
\label{Ssfinal}
S_s &=& - \frac{1}{2} \int \frac{\d^2 {\bf p}}{(2\pi)^2} \, {\bf p}^2 \left[  1 + g^2 G({\bf p}^2) \right] \tilde \phi({\bf p}) \tilde \phi(-{\bf p}) \ ,
\end{eqnarray}
where we have defined
\be
\label{Gdef}
G({\bf p}^2) \equiv \int \frac{\d^{d-2} \overline p}{(2\pi)^{d-2}} \frac{\tilde f(\overline p) \tilde f(-\overline p)}{{\bf p}^2 + \overline p^2} \ ,
\ee
dividing the momentum $p$ up into components ${\bf p}$ parallel and $\overline p$ perpendicular to the defect and where 
\be
\tilde f(\overline p) = \int \d^{d-2} y \, e^{-i \overline p \cdot y} f(y) 
\ee
is the Fourier transform of the smearing function.  Note in the case $d=2$ where $G({\bf p}^2) = {\bf p}^{-2}$, we recover the standard Schwinger model result, with a pole in the propagator ${\bf p}^2 = - g^2$ corresponding to the square of the coupling strength.

We leave the details of the case of $N_s$ scalars as an exercise.  One finds $N_s-1$ free, massless scalars and one scalar with the action (\ref{Ssfinal}) but with $g^2 \to N_s g^2$.  The special scalar is a normalized linear combination $\phi = \frac{1}{\sqrt{N_s}} \sum_i \phi_i$ of the original scalars.

\subsection{Spinors}

For the fermions ${\mathcal L} = {\mathcal L}_f$, we have instead the action
\begin{eqnarray}
\label{spinorS}
S_f &=&  \int \d^d x \, \left(  - \frac{1}{4}  F_{\mu\nu} F^{\mu\nu}  - \frac{1}{2\zeta} (\partial^\mu A_\mu)^2 
+ g A_a(x) J^a({\bf x}) f(y)
\right)
+i   \int \d^2 {\bf x}  \, \bar \psi \slashed{\partial} \psi  \ ,
\end{eqnarray}
where $J^a = \bar \psi \gamma^a \psi$ for Minkowski signature $\gamma$ matrices in $1+1$ dimensions.  Ignoring the gauge fixing term, gauge invariance is preserved for $f(y) = \delta(y)$ but 
destroyed for a more general choice, in contrast to the scalar case analyzed above.

We follow the steps in the scalar case, first integrating out the transverse components of the gauge field $A^i$, yielding
\begin{eqnarray} \label{eq:SfbeforeIntegrationOut}
S_f &=& \int \frac{\d^d p}{(2\pi)^d} \left[ - \frac{1}{2} \left( p^2 \eta_{ab} - \frac{p_a p_b (1-\zeta^{-1})}{1- \frac{p^k p_k}{p^2} (1-\zeta^{-1})} \right) \tilde A^a(p) \tilde A^b(-p) + i g  \tilde A_b(p)  \tilde {\mathcal J}^b(-p) \right] \nonumber \\
&& +i \int \d^2 {\bf x}\bar \psi \slashed{\partial} \psi \ ,
\end{eqnarray}
where we have defined ${\mathcal J}(x) = J({\bf x}) f(y)$.  
And then integrating out $A_a$ to produce 
\begin{eqnarray}
\label{Sftemp}
S_f &=&  \frac{g^2}{2}  \int \frac{\d^2 {\bf p}}{(2\pi)^2}G({\bf p}^2) \tilde J^a ({\bf p}) \tilde J_a (-{\bf p})  + i \int \d^2 {\bf x}\bar \psi \slashed{\partial} \psi
\end{eqnarray}
where $G({\bf p}^2)$ is as defined in (\ref{Gdef}). 

 In contrast to the scalar case, this action is not quadratic and a different solution strategy is required.    
 We first note that (\ref{Sftemp}) can be produced by integrating out a nonlocal Maxwell field $B_a$ from a purely 2d theory:
 \begin{eqnarray}
 \label{Sfequiv}
 S_f &=& \int \frac{\d^2{\bf p}}{(2 \pi)^2} \left[ -\frac{1}{2 G({\bf p}^2)} \tilde B^a({\bf p}) \tilde B_a(-{\bf p}) + g \tilde { J}^a(-{\bf p}) \tilde B_a({\bf p}) \right] + i \int \d^2 {\bf x}\bar \psi \slashed{\partial} \psi \ .
 \end{eqnarray} 
 This partially Fourier transformed action is gauge fixed with $\zeta = 1$.  In position space without the gauge fixing, we may write schematically the action as
 \begin{eqnarray}
 \label{Sfequiv2}
 S_f &=&\int \d^2{\bf x} \left[-\frac{1}{4} W^{ab} \frac{1}{-\partial^2 G(-\partial^2)}W_{ab}  + i\bar \psi \slashed{\partial} \psi  + g \bar \psi \slashed{B} \psi \right] \ ,
 \end{eqnarray} 
 where $W_{ab} = \partial_a B_b - \partial_b B_a$.  This rewriting makes explicit the nonlocal nature of the effective 2d theory.

A little detour follows, following closely the strategy described in \cite{ZinnJustin} for the original Schwinger model.  As a special case of Hodge decomposition in two dimensions, we can re-express the 2d gauge field as
\be
\label{Basub}
B_a = \frac{1}{g} \left( \partial_a \chi + \epsilon_{ab} \partial^b \rho \right)\ ,
\ee
for scalar fields $\rho$ and $\chi$.  To decouple the gauge field from the fermion, we make the following field redefinitions
\be\label{fermionRotations}
\psi = e^{i \chi - i \gamma_5 \rho} \psi' \ , \; \; \; \bar \psi = \bar \psi' e^{-i \chi - i  \gamma_5 \rho} \ .
\ee
Under this transformation, $i\bar \psi \slashed{\partial} \psi  + g \bar \psi \slashed{B} \psi  \to i\bar \psi' \slashed{\partial} \psi'$ while the nonlocal Maxwell term gives rise to a nonlocal scalar kinetic term
\be
S_f = \int \frac{\d^2{\bf p}}{(2 \pi)^2}\frac{{\bf p}^2}{2 g^2 G({\bf p}^2)}  \tilde \rho({\bf p}) \tilde \rho(-{\bf p}) +  i \int \d^2 {\bf x}\bar \psi' \slashed{\partial} \psi'   \ .
\ee  
This field redefinition however has a nontrivial Jacobian under the path integral which introduces an extra free scalar kinetic term for $\rho$, but with the wrong sign and particular normalization:
\be
S_f \to S_f + \frac{1}{2\pi} \int \d^2{\bf x} (\partial_a \rho)^2 \ .
\label{Jacobianfactor}
\ee
Finally, we can trade the free fermion $\psi'$ for a free massless scalar $\Phi$ under bosonization:
\be
 i \int \d^2 {\bf x}\bar \psi' \slashed{\partial} \psi'  \to -\frac{1}{2} \int \d^2 {\bf x} (\partial_a \Phi) (\partial^a \Phi) \ .
\ee
Our final action is then, dropping the gauge mode $\chi$,
\begin{eqnarray}
\label{almostfinal}
S_f &=& \int \frac{\d^2{\bf p}}{(2 \pi)^2} \biggl[
- \frac{{\bf p}^2}{2} \tilde \Phi({\bf p}) \tilde \Phi(-{\bf p}) + \left( \frac{{\bf p}^2}{2 g^2 G({\bf p}^2)} + \frac{{\bf p}^2}{2\pi} \right) \tilde \rho({\bf p}) \tilde \rho(-{\bf p})  
\biggr] \ .
\end{eqnarray}
A ghost-like mode in $\rho$ can be isolated through the redefinition $\Phi \to \Phi + \frac{1}{\sqrt{\pi}} \rho$, leading to the new action
\begin{eqnarray}
S_f &=& \int \frac{\d^2{\bf p}}{(2 \pi)^2} \left[
- \frac{{\bf p}^2}{2} \tilde \Phi({\bf p}) \tilde \Phi(-{\bf p}) + \frac{{\bf p}^2}{2 g^2 G({\bf p}^2)} \tilde \rho({\bf p}) \tilde \rho(-{\bf p})  
- \frac{{\bf p}^2}{\sqrt{\pi}} \tilde \Phi({\bf p}) \tilde \rho(-{\bf p}) \right] \ .
\end{eqnarray}
Finally integrating out $\tilde \rho$ produces
\begin{eqnarray}
S_f &=& - \int \frac{\d^2{\bf p}}{(2 \pi)^2} \, \frac{{\bf p}^2 }{2}\left[ 1 + \frac{g^2}{\pi} G({\bf p}^2) \right] \tilde \Phi({\bf p}) \tilde \Phi(-{\bf p}) 
\label{altSf}
\end{eqnarray}
which matches the form of the action in the scalar case previously (\ref{Ssfinal}), up to the rescaling $g \to g / \sqrt{\pi}$ discussed above associated with matching the normalizations of the two currents under the bosonization rules. 
 Similar to the scalar case, one can also add a $\epsilon^{ab} F_{ab}$ term to the action, which acts to shift the scalar field $\Phi$, and thus can be undone through a field redefinition of $\Phi$.  

We leave the generalization to $N_f$ fermions as an exercise.  Repeating the steps above (and using abelian bosonization), one finds $N_f-1$ decoupled free fermions (or equivalently free scalars) and one free scalar with the action (\ref{altSf}) but with $g^2 \to N_f g^2$. The rescaling of the coupling can be traced to the rescaling of the Jacobian factor in (\ref{Jacobianfactor}) by $N_f$, which in turn happens because the change of variables is carried out on $N_f$ independent fermions.

\section{Effective Propagator}
\label{sec:analysis}

\begin{figure}
\begin{center}
\includegraphics[width=3in]{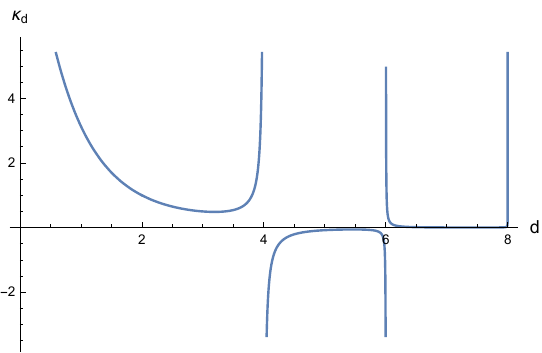}
\end{center}
\caption{
A plot of $\myconst$ vs.\ $d$ where $G({\bf p}^2) = \myconst | {\bf p}|^{d-4}$.
The constant $\myconst$ has a local minimum at $d\approx 3.19$.
\label{fig:coeffplot}
}
\end{figure}

We saw above that both the scalar and spinor models gave rise to the same quadratic action (\ref{Ssfinal}) and (\ref{altSf}) for a scalar degree of freedom, up to the replacement $g^2 \to g^2 / \pi$.  Here we analyze the propagator that follows from the action:
\be
\label{mypropagatorgeneral}
\Pi({\bf p}^2) = \frac{1}{{\bf p}^2 (1 + \alpha G( {\bf p}^2 )) } \ ,
\ee
where $\alpha \equiv g^2$ or $g^2/\pi$ as appropriate.

If $f(y) = \delta(y)$, then the integral (\ref{Gdef}) for $G({\bf p}^2)$ converges in the range $2<d<4$:
\be\label{eq:Gp}
G({\bf p}^2) =  \myconst |{\bf p}|^{d-4}  \mbox{    where    } \myconst \equiv
\frac{\Gamma\left(\frac{4-d}{2}\right)}{(4\pi)^{\frac{d}{2}-1}}\ ,
\ee
assuming ${\bf p}^2>0$.\footnote{%
Curiously, a nearly identical propagator appears in the analysis of a free six dimensional scalar coupled to a non-linear sigma model on a defect in ref.~\cite{Cuomo:2023qvp}.
We have more to say about this coincidence in the Discussion below.
\label{footnote:coincidence}
}
Indeed, thinking of ${\bf p}^2$ as a complex variable, the result can be extended to the entire complex plane away from the negative real axis.  In dimensional regularization, more ambitiously, we may try to make sense of this result not only in the range $2< d < 4$ but for all positive $d \neq 4, 6, 8 \ldots$ where $\myconst$ is finite.  

In the physically relevant case $d=4$, we can extend the analysis through a variety of different regularization schemes.  We can expand in small $\epsilon$, in $4-\epsilon$ dimensions, which yields a logarithmic behavior in the propagator.  With a hard momentum cutoff $|{\bf p}| < \Lambda$, the
function $G({\bf p}^2)$  evaluates to 
\be
\label{Gd4hc}
G({\bf p}^2) = \frac{1}{4\pi} \log \left(1+\frac{\Lambda^2}{\pb^2}\right).
\ee
We also explore the softer Gaussian regulator used in \cite{Gorbar:2001qt}  in appendix \ref{app:regulator}.
In all cases, the result
looks similar to the one loop self energy of the scalar in massless $\phi^4$ theory in 4d or of the photon in 4d QED. The difference is that while the self-energy in these theories resums only a class of diagrams, our propagator is exact given the choice of UV regulator. 

Like the resummed propagator in QED, there is effectively a Landau pole in the UV in this $d=4$ case with a hard cut-off, 
when $1 + g^2 G({\bf p}^2) = 0$ vanishes, or equivalently
\be
{\bf p}^2 = - \frac{\Lambda^2}{1 - e^{-4 \pi /g^2}} \ ,
\ee
which in magnitude is of order or larger than the UV cut-off scale $\Lambda$, and thus cannot be trusted.  Detailed knowledge of the 
UV completion is required to determine exactly what happens at high scale.
Indeed with a different Gaussian cut-off, we will see in appendix \ref{app:regulator} this pole is absent.

The attitude in this analysis is to treat the results in the full range $d>0$  
as data that can help us understand the obviously 
physically relevant cases 
of $d=2$, $3$ and $4$.
We start with the range $0<d<4$, where at large $p$ the propagator looks like that of a free massless scalar in $2d$, with scaling dimension 0.  Small $p$ on the other hand is dominated by the $| {\bf p}|^{d-4}$ term which by dimensional reasoning should give a scalar two-point function with dimension  $\frac{4-d}{2}$.  In other words, we can interpret the coupling of the surface to the Maxwell field as inducing a flow from a free scalar theory in the UV to an effective CFT with a scalar of dimension $\frac{4-d}{2}$ in the IR.  We recast this claim more formally in the language of renormalization and beta functions below.  This reasoning fails for $d=2$, where we get the Schwinger model and a flow from a free theory in the UV to a gapped theory (with mass $m^2 = g^2 $) in the IR and also in $d=4$ where the integral expression for $G({\bf p}^2)$ is not well defined without a regulator.  We will see below that the range $0<d<2$  is problematic for another reason:  the spectral density is negative.  

Physically, the effect of the electromagnetic field on the charged particles should weaken at long distances as $d$ grows.  Coulomb's Law for the potential around a charged particle, after all, has the power law behavior $r^{3-d}$.  The above behavior is largely consistent with this intuition.  Whether Maxwell theory makes sense in $d<2$ is questionable but in $d=2$, where the effect is strong, we get a mass gap  in the IR.  In the range $2<d<4$, the anomalous dimension of the scalar changes along the flow, changing the most (by one) in the limit $d \to 2$ and changing the least (by nothing at all) in the limit $d \to 4$.

It is curious that the generic behavior in this set-up appears to be a flow from one CFT to another.  The lore from studies of critical phenomena is that CFTs require special tuning while gapped phases are generic.  Here the opposite appears to be true.  We get a gapped phase for the carefully tuned value $d=2$ while generically in the range $2<d<4$ the IR is a CFT.  Perhaps this result is a consequence of the quadratic nature of the underlying action, that the theory in some sense is not truly interacting.

If we ambitiously try to analyze the cases where $d>4$, $d \neq 6, 8, \ldots$, taking the result (\ref{mypropagatorgeneral}) combined with (\ref{eq:Gp}) at face value,
then the direction of the renormalization group flow changes.
The coupling is irrelevant and a scalar
of dimension $\Delta = \frac{4-d}{2} < 0$ in the UV naively flows to a scalar of dimension zero in the IR. Of course
the scalar at the UV fixed point is below the unitarity bound and it is far from clear whether this scenario is physically meaningful.
There are other peculiarities associated with the cases $d>4$ that we will discuss in more detail below.

\subsection{Beta functions}

We can perform a conventional RG analysis of the quadratic action obtained from both the bosonic \eqref{Ssfinal} and fermionic \eqref{altSf} theories; this analysis will be reminiscent of the renormalisation of the large-$N$ $\phi^4$ theory. In particular, the coupling flows only due to its engineering dimension and the field renormalisation. We first write the bare action from (\ref{Ssfinal}) or (\ref{altSf}) 
\begin{align}
S_f &= -\int\frac12 \phi_0 (-\partial^2) \phi_0 -\alpha_0 \int \frac12 \phi_0   [-\partial^2 G(-\partial^2)] \phi_0 \ .
\end{align}
Now, we change to the renormalised field and coupling
\be
\phi_0 = Z_\phi \phi, \quad \alpha_0 = Z_\alpha \alpha \mu^{4-d},
\ee
defining the renormalisation scale $\mu$. The propagator of the renormalised field $\phi$ is 
\be
\Pi(\pb^2) = \frac{1}{\pb^2 Z_\phi^2 (1+ Z_\alpha \alpha \mu^{4-d} G(\pb^2))}.
\ee
For $d\neq 2$ (where we would find a mass term)\footnote{%
 The case $d=2$ requires special treatment as the renormalization condition (\ref{renormcond}) is no longer appropriate. The massive propagator $1 / (p^2 + \alpha_0)$ does not require a $Z_\phi$ factor, and the mass flows purely according to its engineering dimension.
} 
we impose the renormalisation condition
\begin{align}
\label{renormcond}
&\Pi(\mu^2) = \frac{1}{\mu^2}\\
\implies & Z_\phi^2 (1+Z_\alpha \alpha \mu^{4-d} G(\mu^2))=1.
\end{align}
This cancels the divergence; we now have at this point an arbitrary choice of $Z_\alpha$, corresponding to our freedom of reparametrisation for $\alpha$. We choose to take $Z_\alpha Z_\phi^2 =1$, in which case we find
\begin{align}
    Z_\phi^2 = 1-\alpha \mu^{4-d} G(\mu^2) \ .
\end{align}
The beta function $\beta_\alpha \equiv d \alpha / d \log \mu$ can be deduced from the scale independence of $\alpha_0$:
\be
\beta_\alpha
= -\alpha(4-d - 2\gamma_\phi), \label{eq:arbGbeta}
\ee
where in the last equality we have written $\beta_\alpha$ in terms of the anomalous dimension for $\phi$,
\be
\gamma_\phi \equiv \dv{\log Z_\phi}{\log\mu} \ .
\label{eq:arbGgamma}
\ee
Evaluating the beta function for $d \neq 2, 4$ with \eqref{eq:Gp}, we find
\begin{align}
       \beta_\alpha = -(4-d) \alpha(1-\alpha \myconst) \ ,\label{eq:DREGbeta}
\end{align}
precisely identical to the large-$N$ beta function of $g\phi^4$ theory, up to the identifications $g \leftrightarrow \alpha$ and $g_\star \leftrightarrow \myconst^{-1}$ ((18.56) of \cite{ZinnJustin}). 
The coincidence is presumably due to the similarities of our action with the action of the auxiliary field in that theory.  Both theories have a one-loop-exact structure that is marginally irrelevant for $d=4$. Just as in $\phi^4$ theory, in generic $d$ we find two fixed points, at $0$ and $\myconst^{-1}$, that collide as $d \to 4$, where we find the beta function $\beta_\alpha= \tfrac{1}{2\pi} \alpha^2$. The anomalous dimension of $\phi$ is 
\be
\gamma_\phi = \beta_\alpha \dv{\log Z_\phi}{\alpha} = \frac{4-d}{2} \alpha \myconst \ , \label{eq:DREGgamma}
\ee
which indeed interpolates between the free field in the UV and the scalar of dimension $\Delta_\phi = 0 + \gamma_\phi = \frac{4-d}{2}$ in the IR. Other scheme choices are a non-linear redefinition of the coupling $\tilde \alpha(\alpha)$. Take, for example, $Z_\alpha =1$; we find for $d \neq 2,4$
\begin{align}
    \beta_\alpha = -(4-d) \alpha, \quad
    \gamma_\phi = \frac{4-d}{2} \frac{\alpha \myconst}{1+ \alpha \myconst}  \ .
    \end{align}
Thus for $d\neq 4$ the fixed points at $\alpha=0$ and $\alpha=\infty$ reproduce the correct IR and UV anomalous dimensions.

Using a hard cut-off in $d=4$ instead of dimensional regularization,
in our first scheme where $Z_
\alpha Z_\phi^2 = 1$, the beta function and anomalous dimension evaluate to
\be
\beta_\alpha = \frac{\alpha^2}{2\pi} \frac{\Lambda^2}{\Lambda^2 + \mu^2}, \quad \gamma_\phi = \half \frac{\alpha}{2\pi}\frac{\Lambda^2}{\Lambda^2 + \mu^2},
\ee
which are identical in the $\Lambda \to \infty$ limit to \eqref{eq:DREGbeta} and \eqref{eq:DREGgamma} evaluated at $d=4$.

\subsection{Spectral Decomposition}

We analyze the spectral properties of this propagator (\ref{mypropagatorgeneral}).  Employing (\ref{eq:Gp}), the exponent $d-4$ requires in general a branch cut.  Placing the branch cut along the negative $\nu = {\bf p}^2$ axis, we can write the spectral density $\chi(\nu)$ as the discontinuity of 
$\Pi(\nu)$ across the branch cut:
\begin{eqnarray}
\chi(\nu) &\equiv& \lim_{\epsilon \to 0^+} \frac{1}{2 i} \left(\Pi(\nu - i \epsilon) - \Pi(\nu + i \epsilon) \right) \nonumber \\
&=& \lim_{\epsilon \to 0^+} \operatorname{Im}(\Pi(\nu - i \epsilon)) \ .
\end{eqnarray}
The full propagator can then be reconstructed from the spectral density by means of a contour integral:
\be
\label{Pitochi}
\Pi(\nu) = \frac{1}{\pi} \int_0^{-\infty}
\frac{\chi(\omega)}{\omega - \nu} d\omega \ .
\ee
The validity of this relation requires that $\Pi(\nu)$ has no additional singularities elsewhere in the plane and also that $d<4$ so that the portion of the contour near the origin can be neglected.

The spectral density $\chi(\nu)$ has some notable features.  For $d<2$, the spectral density is negative along the negative ${\bf p}^2$ axis, indicating a pathology in the theory.
The propagator $\Pi({\bf p}^2)$ itself has poles, above and below the negative ${\bf p}^2$ axis, cousins of the mass pole in the standard case $d=2$.  These poles on the physical sheet spoil the contour integral argument and require extra contributions be added to (\ref{Pitochi}).

\begin{figure}
\begin{center}
    a)
\includegraphics[width=2.5in]{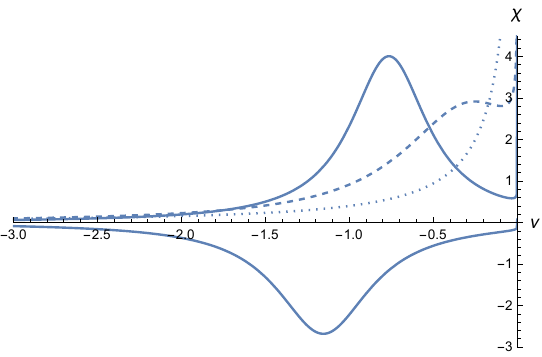}
    b)
    \includegraphics[width=2.5in]{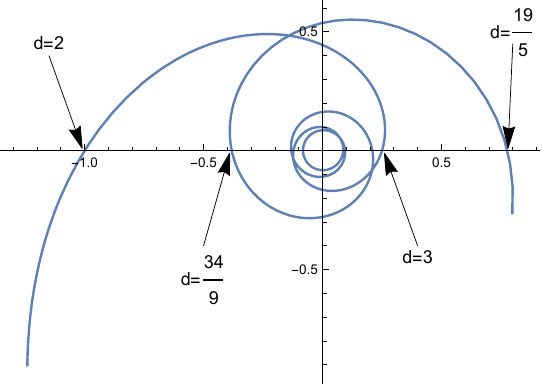}
    \end{center}
\caption{a) The spectral density as a function of $\nu = - {\bf p}^2$ for $g=1$ and $d = 1.8$ (solid curve with $\chi<0$), $d=2.2$ (solid curve with $\chi>0$), $d=2.5$ (dashed curve) and $d=3$ (dotted curve). b) Location of one of the poles in the complex ${\bf p}^2$ plane as a function of $d$ for $g=1$.  There is a second branch point which is the complex conjugate of this one, spiraling in the opposite direction.  As $d$ increases from 2, the branch point moves further and further from the physical sheet.
The pole is closest to the origin when $d\approx 3.58$.
\label{fig:branchpoint}}
\end{figure}

In contrast, for $2<d<4$, the spectral density is positive.  For $d \gtrsim 2$, one can still see a sharp peak close to $\nu = -g^2$ -- continuously related to the mass pole for $d=2$. 
However, for $d>2$, these mass poles have moved off the physical sheet into neighboring sheets and thus allow for the contour integral argument behind (\ref{Pitochi}).  As $d$ increases further toward $4$, these poles spiral around the origin, moving farther and farther away from the physical sheet.
One of these poles is shown in figure \ref{fig:branchpoint} as a function of $d$.

Extending the analysis to $d=4$ with a hard cut-off, the spectral density reassuringly is positive in the range $-\Lambda^2 < \nu < 0$.  For smaller values $\nu < -\Lambda^2$, the spectral density vanishes away from the Landau pole.  (With a Gaussian regulator, the spectral density is also positive.)

\begin{figure}
\begin{center}
    a) \includegraphics[width=0.9in]{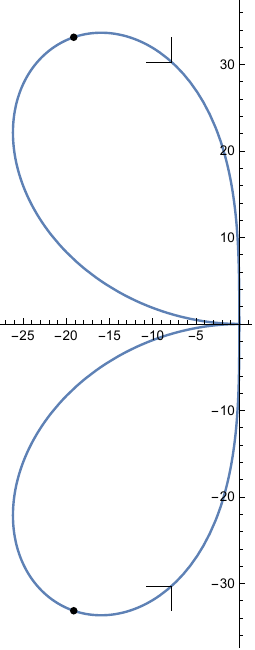}
    b) \includegraphics[width=2in]{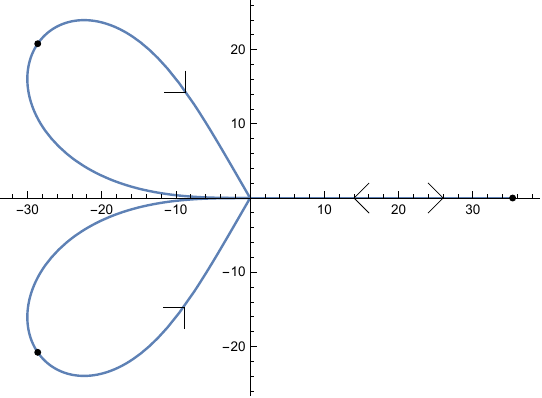}
    c) \includegraphics[width=2in]{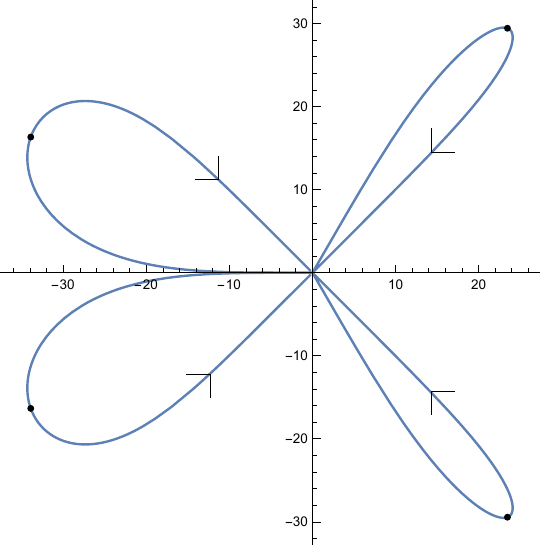}
\end{center}
\caption{
The location of poles on the physical ${\bf p}^2$ sheet in units where $g=1$ in the ranges a) $6<d<8$, b) $8<d<10$, and c) $10<d<12$.  The marked points are the locations of the poles for $d=7$, 9 and 11 respectively.  The arrows indicate the motion of the poles as $d$ increases. When $4<d<6$, there is a single tachyon along the positive ${\bf p}^2$ real axis. \label{fig:badpoles}
}
\end{figure}

Ignoring for the moment
the IR convergence condition that $d<4$ for the validity of the contour argument, we note in passing that 
$\chi(\nu)$ is positive along the negative real axis for $d>4$, $d \neq 6, 8, 10, \ldots$. 
However, the propagator still has a number of peculiarities associated with its pole structure in this range. 
For $4n < d < 4n+2$, $n=1, 2, \ldots$, the $\myconst$ coefficient is negative and there is always a tachyon, i.e.\ a pole at positive values of ${\bf p}^2$. 
While there are no poles for positive ${\bf p}^2$ in the range $4n+2 < d < 4n+4$, starting with $n=2$ there will be pairs of complex conjugate poles with positive real part.  Indeed, the propagator in the range $2n+2 < d < 2n+4$ will generically have $n$ poles on the physical sheet in addition to a branch cut along the negative real axis.  
Given $d>4$, the range $6 < d < 8$ is perhaps the best behaved, having a pair of complex conjugate poles with negative real part.  One may speculate what this result may or may not mean for this type of surface defect in 7 dimensional Maxwell theory.  Figure \ref{fig:badpoles} shows the location of poles on the physical sheet in the range $6<d<12$.

We make one last passing remark before moving on to the behavior of the Green's function in coordinate space.  In the range $2\leq d<4$, the dimensions 
\be
d = \frac{ 2(2q-1)}{q} \ ,
\ee
where $q=1, 2, 3, \ldots$, play a special role.  The propagator in this case can be written in the form
\be
\Pi(\nu) = \left(\frac{-1}{\myconst g^2}\right)^{q-1} \frac{1}{\nu^{\frac{1}{q}} + \myconst g^2} -  \sum_{n=1}^{q-1} \left( \frac{-1}{\myconst g^2} \right)^n \frac{1}{ \nu^{1 - \frac{n}{q}} }  \ .
\ee
Also, at these special values, there are poles which cross the real axis although not on the physical sheet.

\subsection{The Fourier Transform and Special Cases}

We analyze the Fourier transformed (Euclidean) real space propagator
\begin{eqnarray}
G({\bf x}) &=& \int \frac{\d^2{\bf p}}{(2\pi)^2} \Pi({\bf p})e^{i {\bf x} \cdot {\bf p}}  
= \frac{1}{2 \pi} \int_0^\infty \frac{J_0(p |{\bf x}|)  \, \d p}{p (1 + 
(\mu / p)^{4-d})} \ , \nonumber
\end{eqnarray}
where $\mu^{4-d} \equiv g^2 \myconst$.  
Quite generally, we find that at small $|{\bf x}|$, the propagator takes the approximate form
\be
G({\bf x}) = -\frac{1}{2\pi} \left( \gamma + \log \frac{\mu |{\bf x}|}{2} \right) + O(|{\bf x}|) \ ,
\ee
while at large $|{\bf x}|$, assuming $2<d<4$, we get instead
\be
G({\bf x}) \sim \frac{2^{2-d} \Gamma \left(2- \frac{d}{2}\right)}{\pi \Gamma \left(\frac{d}{2}-1 \right)}(\mu |{\bf x}|)^{d-4} \ , 
\ee
consistent with identifying the scalar as a free field of dimension zero in the UV and a generalized free field of dimension $\frac{4-d}{2}$ in the IR.

The integral can be evaluated exactly in special cases.   In $d=2$, we get the standard modified Bessel function
result $G({\bf x}) =  K_0(g |{\bf x}|)/2\pi$.  At small $|{\bf x}|$, the behavior is logarithmic, while at large $|{\bf x}|$ it falls exponentially, $e^{-g |{\bf x}|} / \sqrt{g |{\bf x}|}$, interpolating between a free and a gapped theory.  
In $d=3$, we get instead
\be
G({\bf x})  = \frac{1}{4} ({\bf H}_0( g^2 |{\bf x}|/2)  - N_0(g^2 |{\bf x}|/2) )  \ ,
\ee
where ${\bf H}_\nu$ is the Struve $H$ function and $N_\nu$ is a Bessel function of the second kind.
This propagator interpolates between $\log |{\bf x}|$ at small $|{\bf x}|$ and $1 / \pi g^2 |{\bf x}|$ at large
$|{\bf x}|$, i.e.\ between a scalar of dimension zero in the UV and a scalar of dimension half in the IR.  
Because of the locations of the additional poles  in the propagator, on the positive ${\bf p}^2$ axis but not on the physical sheet, 
there is no exponentially damped contribution to the real space propagator.

For $d=5/2$, we find a Meijer $G$ function which interpolates between $\log |{\bf x}|$ at small $|{\bf x}|$ and
$|{\bf x}|^{-3/2}$ at large $|{\bf x}|$. However, in this intermediate case, there are also two exponentially falling and complex conjugate terms
with a $|{\bf x}|^{-1/2}$ coefficient:
\be
G({\bf x}) \sim c_1 |{\bf x}|^{-3/2} + \left(c_2 e^{(-1)^{5/6} \mu |{\bf x}|} + c_2^* e^{-(-1)^{1/6} \mu |{\bf x}|}\right) \frac{1}{|{\bf x}|^{1/2}} + \ldots
\ee
where $c_1$ and $c_2$ are computable.

Unfortunately, the Fourier transform for the regulated $d=4$ case is IR divergent, blowing up like $\log \log {\bf p}^2$ in the small momentum limit.  
This IR divergence is different from what is found in $\phi^4$ theory or QED in 4d, where the $d^4 p$ measure factor is enough to regulate the small $p$ behavior.

\subsection{Monotonicity}

We have the claim that in the region $2<d<4$, the theory flows from a free scalar of dimension $\Delta_{\rm UV} = 0$ in the UV
to a generalized free field of dimension $\Delta_{\rm IR} = \frac{4-d}{2}$ in the IR.  There is a theorem that 2d CFTs should obey a monotonicity property under RG flow, that the UV central charge $c_{\rm UV}$ should be larger than the IR central charge $c_{\rm IR}$
\cite{Zamolodchikov:1986gt}. 
As our nonlocal 2d theory does not have a stress tensor, it is problematic to define a central charge $c$, which is usually thought of as a coefficient in the two-point function of the stress tensor or alternately as an anomalous contribution to the trace of the stress tensor on a curved background.  There is however still a way of proceeding in our case.  The central charge also shows up as  a coefficient of a log divergence to the partition function on spheres.  We can thus attempt to compute the partition function of a generalized free field on a two-sphere.

The sphere partition function of a generalized free field is known.
The claim 
is that
\be
W(0) - W(\Delta)  \sim 
\left[ 1 - (1-\Delta)^3 \right] \log \Lambda \ .
\label{Wdiff}
\ee
As $0<\Delta = \frac{4-d}{2} < 1$, the difference above is positive, in agreement with the $c$-theorem adapted to this case that lacks a stress tensor. Moreover, the ``distance'' between the IR and UV fixed points, measured as the size of this coefficient, decreases as $d$ increases from 2 to 4, consistent with the fact that the effective strength of the electromagnetic field decreases as $d$ increases.
We rederive (\ref{Wdiff}) in appendix \ref{app:sphereW}, where we also point the reader to some relevant literature.

\section{Wilson and Polyakov Loops}
\label{sec:loopandcondensate}

In the introduction, we mentioned briefly how Wilson loops in the Schwinger model share some qualitative similarities with Wilson loops in QCD.  Here we explore what happens to Wilson and Polyakov loops as a function of the space-time dimension of the photon.

\subsection{Wilson Loops}

The Wilson loop is often used as a diagnostic of confinement, and in the context of this Schwinger model, we can see exactly how the presence of matter affects its expectation value.  Relevant computations when $d=2$ 
can be found in older literature, for example \cite{Sachs:1991en,Smilga:1992yp}.  In the present context, we get to  explore how the nonlocality of the Maxwell field affects the Wilson loop, i.e.\ we get to explore the behavior as a function of $d$.  

For an Abelian gauge field, we define our Wilson loop operator to be
\be
W(C) =  \exp \left( i e\oint_C B \cdot  \d {\bf x}  \right)\ ,
\ee
where the (dimensionful) charge $e$ does not necessarily need to be related to the original coupling $g$.  
We have restricted the curve $C$ to lie in the 2d surface that contains our matter fields.  More precisely, we are evaluating this Wilson loop in the theory after integrating out the directions transverse to the surface defect, for the effective 2d photon $B$ of \eqref{Sfequiv}.  In this context, where we know $\langle B_a ({\bf x}) B_b ({\bf x}') \rangle$ exactly and all connected higher points functions of the gauge field vanish, we can write the expectation value of the Wilson loop in the form
\be
\langle W(C) \rangle = \exp \left( - \frac{e^2}{2} \oint_C \oint_C \langle B_a({\bf x}) B_b({\bf x}') \rangle \d x^a \d x'^{b} \right) \ .
\ee
Through Stokes' Theorem (or really Green's Theorem given the 2d nature of the problem), we replace this double loop integral
by a double area integral over the field strength:
\be
\langle W(C) \rangle = \exp \left( - \frac{ e^2}{2} \int_{D \times D}  \langle W_{01}({\bf x}) W_{01}({\bf x}') \rangle \d^2{\bf x} \, \d^2 {\bf x}'  \right) \ .
\ee

As the next step, we trade the field strength $W_{01}$ for the scalar field $\rho$ introduced above, using (\ref{Basub}):
\be
W_{01} = -\frac{1}{g} \partial_a \partial^a \rho \ ,
\ee
and use the two-point function for $\rho$ that can be deduced from (\ref{almostfinal}).  
One finds
\begin{eqnarray}\label{eq:F01-2pt}
 \langle W_{01}({\bf x}) W_{01}({\bf x}') \rangle &=& \int \frac{\d^2 {\bf p}}{(2\pi)^2} \frac{{\bf p}^2  G({\bf p}^2)}{1 + \frac{g^2}{\pi}  G({\bf p}^2)} e^{i {\bf p} \cdot ( {\bf x} -{\bf x}' )} \ , 
 \label{EEexpectation}
\end{eqnarray}
recalling the result (\ref{eq:Gp}).

For simplicity in what follows, we take the curve $C$ to be a circle with radius $R$.   
To evaluate the Wilson loop expectation value, it is helpful to perform the spatial integrals first.  
The result is
 \be
 \label{circularloop}
 \langle W(C) \rangle = \exp \left( - \frac{\pi^2 e^2}{g^2} 
 (\mu R)^{4-d}
\int_0^\infty \frac{p^{d-3} J_1 ( p)^2}{1 + (\mu R / p)^{4-d}} \, \d p 
 \right) \ ,
 \ee
 where we have defined the coupling parameter
 \be
 \mu^{4-d} \equiv \frac{g^2 }{\pi} \myconst \ .
 \ee

 Let us focus first on the case
 $g=0$, i.e.\ free Maxwell theory.
 The Bessel function integral can be done with the help of (6.574) from \cite{gradshteyn2014table}.\footnote{%
 \[
 \int_0^\infty p^\alpha J_1(p)^2 \d p = \frac{2^\alpha \Gamma(-\alpha) \Gamma\left(\frac{3+\alpha}{2}\right)}
 {\Gamma \left(\frac{1-\alpha}{2} \right)^2 \Gamma \left( \frac{3-\alpha}{2} \right)} \ ,  \; \; \; -3 < \alpha < 0 \ .
 \]
 }
   One finds
 \be
 \langle W(C) \rangle = 
 \exp \left(
 -\frac{e^2 R^{4-d}}{2}  \frac{\pi^{2-\frac{d}{2}} \Gamma(3-d) \Gamma \left(\frac{d}{2}\right)}{\Gamma \left(2-\frac{d}{2}\right)
 \Gamma \left(3 - \frac{d}{2} \right)} 
 \right) \ .
 \ee
The  area law  is recovered in $d=2$, consistent with the linear dependence of the potential in this dimension.  As $d$ increases and the Maxwell field gets weaker at long distances, one finds a generic power law behavior $R^{4-d}$.  Once $d \geq 3$, the Wilson loop diverges and requires a UV regulator.  Returning to the original picture where we are integrating over the perimeter $C$ instead of the bulk $D$, the UV regulator should come into effect when the points ${\bf x}$ and ${\bf x}'$ approach each other on the circle, leading to a perimeter law with a coefficient that depends on the details of the regulator. 

Now consider the case of nonzero $\mu$. The integral continues to diverge for $d \geq 3$, which we interpret vis-a-vis a UV cut-off as perimeter law behavior.  For certain specific choices of dimension in the range $ d < 3$, the integral can be done exactly.  For example, when $d=2$, we find
\be
\int_0^\infty \frac{p J_1(p)^2}{p^2+(\mu R)^2}  \d p = I_1 (\mu R) K_1(\mu R) \ .
\ee
At small coupling, one finds a logarithmic correction to the area law,
\be
I_1 (\mu R) K_1(\mu R) = \frac{1}{2} + \frac{1}{4} (\mu  R)^2 \log \mu R + \ldots \ .
\ee
At large  coupling on the other hand, the area law shifts to a perimeter law,
\be
I_1 (\mu R) K_1(\mu R) \sim \frac{1}{2\mu R}  \ ,
\ee
where the expectation value of the Wilson loop scales with $R$ instead of $R^2$.  
This result is reminiscent of certain exact results for circular Wilson loops in supersymmetric gauge theories where Bessel functions also arise \cite{Erickson:2000af}. 

Indeed, quite generally we find that the large coupling limit of the Wilson loop expectation value leads to perimeter law behavior although it is not obvious from the form of (\ref{circularloop}).
To estimate the behavior of this integral, we break the integration range into two pieces, one from $p=0$ to the first maximum of the Bessel function, the second from the first maximum to $p\to \infty$.  
In the first interval, we estimate $J_1(p) \lesssim \frac{p^2}{4}$.  In the second interval, we instead take $J_1(p)^2 \lesssim \frac{2}{\pi p}$.
We focus on the dimension range $2 \leq d < 3$.
In the first interval, the integral is dominated by the $(\mu R/p)^{d-4}$ term in the denominator, which will lead to constant scaling of the expectation value with $R$.  
The large $p$ region is more interesting.  The integral has a contribution which is bounded above by a term that scales as $R^{d-3}$, which if close to saturated would again lead to perimeter law scaling and dominate the contribution from the region close to $p=0$. 
To get a more quantitative estimate, we replace $J_1(p)^2$ with its average large-$p$ value $\frac{1}{\pi p}$ instead of the upper bound $\frac{2}{\pi p}$.  The integral can be done exactly and yields
\be
\int_0^\infty \frac{p^{d-3} J_1 ( p)^2}{1 + (\mu R /p)^{4-d}} \, \d p \sim 
\frac{(\mu R)^{d-3}}{(4-d) \sin \left( \frac{\pi}{4-d} \right)}
\ee
in the large $\mu$ limit.  This $R^{d-3}$ scaling cancels with the $R^{4-d}$ prefactor in the Wilson loop expectation value to give perimeter scaling in the large $R$ limit.

\subsection{Polyakov loops}

The Polyakov loop operator is often used as an order parameter for confinement transitions.  To define it, we imagine our surface defect is a cylinder with circumference $\beta$, which we can also interpret as one divided by the temperature.  
The Polyakov loop operator is then the specific case of a Wilson operator where
\be
P_e(x) = \exp \left(i e \int_0^\beta d\tau B_0(\tau, x)\right) \ .
\ee

We are interested in particular in the expectation value of a pair of Polyakov operators separated by a distance $L$. This is just a Gaussian integral, and so following the same steps as in the Wilson loop case above, we can write this expectation value in terms of the two point function of the electric field integrated over the cylinder separating the two loops:
\be
\langle P_e(0) P_{-e} (L) \rangle 
=
\exp \left( -\frac{e^2}{2} \int_{D \times D} \langle W_{01}({\bf x}) W_{01}({\bf x}') \rangle \d^2 {\bf x} \, \d^2 {\bf x}'\right)
\ee

Moving to Fourier space, we keep an integral over the spatial momentum $p$, but periodicity means the integral over the time-like momentum is traded for a sum over Matsubara frequencies.  Performing the two time integrals first, only the $\omega = 0$ Matsubara frequency survives in the sum.  Our expectation value becomes
\be
\langle P_e(0) P_{-e} (L) \rangle 
=
\exp \left( -\beta e^2
\int \frac{\d p}{2\pi} \frac{G(p^2)}{1 + \frac{g^2}{\pi} G(p^2)} (1-\cos p L) \right) \ .
\ee

The result in the case $d=2$ is particularly simple
\be
\langle P_e(0) P_{-e} (L) \rangle 
=
\exp \left(-
\frac{\beta e^2}{2 \mu} \left( 1 - e^{-\mu L} \right)
\right) \ ,
\ee
which agrees with an old result from \cite{Gross:1995bp}.
There is a transition from area law at short distances to something that scales with $\beta$ up to exponentially small corrections at long distances.  

We can also now analyze the general $d$ case, provided $d<3$.  To get the small distance behavior, we drop the coupling dependence:
\be
\langle P_e(0) P_{-e} (L) \rangle  \sim
\exp \left( -\beta L^{3-d} e^2 \frac{ \myconst \sec \frac{\pi d}{2}}{ 2\Gamma(4-d)} 
\right) \ .
\ee  
In this case, we should simply recover the evaluation of a Wilson loop in pure Maxwell theory, as the fact that the loop is living on a defect is irrelevant.

At large separation, we can similar to before estimate the integral by replacing $1 - \cos p L$ by its average value 1/2:
\be
\langle P_e(0) P_{-e} (L) \rangle  \sim
\exp \left( -
\frac{\pi \beta e^2}{ g^2} 
\frac{\mu}{  (4-d) \sin \frac{\pi}{4-d} }
\right) \ ,
\ee
which scales linearly with $\beta$.

For $d\ge 3$ we expect a UV divergence, which when UV-regulated will yield a perimeter law, $\expval{P_e(0)P_{-e}(L)} \sim \exp(-\text{const} \times \beta)$. These cases include $d=4$.

\section{Discussion}
\label{sec:discussion}

The original motivation for this work was the question what happens when  2d massless fermions are coupled to a 4d Maxwell field, and we should try to provide an answer.  There appear to be strong parallels with the behavior of 
 4d QED or $\phi^4$ theory, where the interaction is marginally irrelevant, flowing to zero in the IR and becoming strong in the UV.  There can even be Landau pole, depending on the choice of UV regularization. 
One significant difference, however, appears to be the IR behavior.  While the Fourier transforms of the propagator in $\phi^4$ field theory should be finite in the IR because of the $\d^4 p$ measure factor,
for our theory, the scalar propagator has an IR divergence, despite the vanishing of the effective coupling.
Moving away to $d=4 -\epsilon$ dimensions, similarities persist.  One finds in all cases a renormalization group flow between a free theory in the UV and a theory with anomalous dimensions in the IR.  As $\epsilon \to 0$, these two fixed points merge.

Refs.\ \cite{Cuomo:2021kfm,Cuomo:2022xgw,Cuomo:2023qvp} found similar ``runaway'' flows in related defect systems, the last two for line defects and the first for surfaces.  The model in ref.\ \cite{Cuomo:2023qvp} in particular shares many similarities with the model here as already noted in footnote \ref{footnote:coincidence}.  
In that work, the authors study a 
$h \int \d^2 x \, \Phi \cdot n$ coupling between free scalars in the bulk and a non-linear sigma model (NLSM) on a surface defect.  Here $\Phi_i$ is one of $N$ free scalars and $n_i$ is one of the $N$ NLSM fields with the constraint $n^2 = 1$.  Enumerating some of the gross similarities between these models, their bulk scalar field $\Phi$ plays a similar role to our Maxwell field $F_{\mu\nu}$ while their NLSM field $n$ is our scalar $\phi$.  
While they have a $h \, n \cdot \Phi$ coupling, we have a $g F_{01} \phi$ coupling.  Their relevant coupling $h$ becomes marginal in the $d\to 6$ limit, while  our relevant coupling $g$ becomes marginal in the $d \to 4$ limit.  
Something very similar to our propagator (\ref{mypropagatorgeneral}) appears as well.  
Assuming spontaneous symmetry breaking of the $O(N)$ symmetry, the Goldstone bosons have the momentum space propagator (\ref{mypropagatorgeneral}) with one small adjustment: the $\myconst |{\bf p}|^{d-4}$ term in the denominator becomes $\myconst |{\bf p}|^{d-6}$.  As a result of that shift, the interesting range of behavior is $4 \leq d \leq 6$ for them while it is $2 \leq d \leq 4$ for us. 
Taking then 
 $d \to 6$ in their case and $d \to 4$ in ours, a divergence in $\myconst$ indicates that the momentum space propagator should be rewritten in terms of a logarithm of the momentum, leading to similar RG flows.

There are some important differences however between these models.  When $d<6$, the coupling $h \Phi \cdot n$ leads to a one-point function for $\Phi$ in the bulk whose form in general is not consistent with conformal symmetry, as it is proportional to the dimensionful $h$.  For us, on the other hand, a gradient of $\phi$, i.e.\ a charge density or current, is required to source a one point function for $F_{\mu\nu}$.  Thus despite the superficial similarity in the bulk-surface coupling, no one-point functions for $F_{\mu\nu}$ are sourced in our case.  
It is conceivable however that a one point function for the stress tensor $\langle T_{\mu\nu} \rangle$ is sourced with a strength proportional to $g$.  
Thus the full theory, including the bulk, may lack conformal invariance even if the defect theory, having integrated out the bulk, preserves it.
It would be interesting to compute $\langle T_{\mu\nu} \rangle$ for our theory in the future.

 There are many other possible directions this work could be extended.  We have focused on the dynamics of the effective scalar degree of freedom, and been largely blind to dynamics specific to the fermions or the $d$-dimensional photons.  It would be interesting to explore physics associated with their correlation functions. The Schwinger model is known to exhibit richer behavior when mass terms for the fermions are turned on.  See for example \cite{Dempsey:2023gib} for recent work on the subject.  It would be interesting to explore what happens in this non-local case, for $d>2$, as a function of fermion mass and $N_f$.  
 Another direction to explore are the constraints of generalized symmetries.  There was a sharp difference in the IR behavior between the cases $d=2$, $2<d<4$, and $d=4$.  Could generalized symmetries help explain this difference, in the style of refs.\ \cite{Delmastro:2022prj,Misumi:2019dwq}? 

\section*{Acknowledgments}
We would like to thank Ofer Aharony, 
Dio Anninos, Gabriel Cuomo, Nadav Drukker, Alan Rios Fukelman, Igor Klebanov, Zohar Komargodski, 
Steve Simon, Shivaji Sondhi, and Andy Stergiou for discussion.
C.H.\ thanks the Oxford University Physics Department for hospitality, where much of this work was prepared.
This work was partially supported by a Wolfson Fellowship from the Royal Society and by the
U.K. Science and Technology Facilities Council Grant ST/P000258/1. L.F.T.\ is supported by a Dalitz Scholarship from the University of Oxford and Wadham College.

\appendix

\section{Conventions}
\label{app:conventions}

Throughout, we have a two dimensional surface defect immersed in a spacetime of dimension $d$.
Any generic point in spacetime is written as $x = ({\bf x}, y) \in {\mathbb R}^{1,d-1}$ where ${\bf x} \in {\mathbb R}^{1,1}$ and 
$y \in {\mathbb R}^{d-2}$.  A point along the surface defect has coordinates $({\bf x}, 0)$.  Greek indices $\mu,\nu, \lambda$ run from $0, \ldots, d-1$.  We reserve Latin indices $a, b$ to index ${\bf x}$ and $i,j$ to index $y$.   Our Minkowski metrix has mostly plus signature $\eta_{\mu\nu} = (-,+, \ldots +)$ which is restricted to $\eta_{ab} = (-,+)$ along the surface defect.  Our conventions for the Levi-Civita tensors are that $\epsilon^{01 \cdots d-1} = 1$ and $\epsilon^{01} = 1$, and indices are raised with the Minkowski tensors $\eta_{\mu\nu}$ or $\eta_{ab}$.  
We will make heavy use of Fourier transforms for which we use the standard conventions
\be
f(x) = \int \frac{d^d p}{(2\pi)^d} e^{i p \cdot x} \tilde f(p) \ , \; \; \;
\tilde f(p) \equiv \int d^d x \, e^{-i p \cdot x} f(x) \ .
\ee

We require spinors embedded on the surface defect. For the Clifford algebra, our conventions are that
\be
\{ \gamma^a, \gamma^b \} = 2 \eta^{ab} \ .
\ee
When necessary, a convenient real basis can be constructed from the Pauli matrices
\be
\gamma^0 = i \sigma^2 \ , \; \; \; \gamma^1 = \sigma^1 \ , \; \; \; \gamma_5 = \gamma^0 \gamma^1 = \sigma^3 \ .
\ee
The conjugate spinors are defined as $\bar \psi = \psi^\dagger \gamma_0$.  Complex conjugation involves a swap in the order
$(\psi \chi)^* \equiv \chi^* \psi^*$. 
With these conventions,
\be
(\bar \psi \psi)^* = - \bar \psi \psi \ , \; \; \;
(\bar \psi \gamma^a \psi)^* = \bar \psi \gamma^a \psi \ , \; \; \;
\left( \int_{{\mathbb R}^{1,1}} d^2 {\bf x} \bar \psi \slashed{\partial} \psi \right)^* = - \int_{{\mathbb R}^{1,1}} d^2 {\bf x} \bar \psi \slashed{\partial} \psi  \ ,
\ee
which means a kinetic term in the Lagrangian must involve a factor of $i$ while the coupling to the gauge field will not.

\section{A Gaussian Regulator}
\label{app:regulator}

While in the main text, we used a hard momentum cut-off $|\overline p| < \Lambda$ to regulate the integral
(\ref{Gdef}) in the case $d=4$, 
here we explore the Gaussian regulator introduced 
in ref.\ \cite{Gorbar:2001qt}:
\be
\tilde f(\overline p) = e^{-{\overline p}^2/2 \Lambda^2} \ .
\ee
We want to test the extent to which the results we saw with a hard cut-off -- a mass pole of order the cut-off and positive spectral density -- depend on our regularization scheme. 
Note that this regulator does not respect gauge invariance for the spinor model.  It would be interesting to explore a regulator that does, for example swapping the line for a small tube of some radius, along which the charged particles are constrained to travel.  

At any rate, for this Gaussian regulator
\begin{align}
G({\bf p}^2) = \int \frac{\d^{d-2} \overline p}{(2\pi)^{d-2}} \frac{e^{- \overline p^2 /  \Lambda^2}}{{\bf p}^2 + \overline p^2} =
\frac{\Lambda^{d-4}}{(4 \pi)^{\frac{d-2}{2}}} e^k E_{\frac{d-2}{2}}(k) \ ,
\end{align}
where we have defined the ratio $k \equiv {\bf p}^2 / \Lambda^2$.    The result is written in terms of the exponential integral
\[
E_n(z) \equiv \int_1^\infty \frac{\d t}{t^n}e^{-z t} \ ,
\]
which has a branch cut discontinuity along the negative real axis in the complex $z$-plane.
To explore the pole structure of the resulting propagator, the analytic continuation formula 
(see 8.19.18 of \cite{NIST:DLMF})
\be
E_p(z e^{2 m \pi i}) = \frac{2 \pi i e^{m p \pi i} \sin(m p \pi)}{\Gamma(p) \sin(p \pi)} z^{p-1} + E_p(z) \ 
\ee
is useful.
We find in fact that with this regulator, the model has no poles at all on the physical sheet for $d=4$.
However, there is a peak in the (positive) spectral density along the negative real ${\bf p}^2$ axis, at
about the scale of the cut-off $\Lambda$.

\section{Sphere Partition Functions}
\label{app:sphereW}

In Fourier space in the interesting range $2<d<4$, 
the kinetic term for the remaining scalar degree of freedom in our theory 
shifts from ${\bf p}^2$ at small distance to ${\bf p}^{d-2}$ at large distance.  
Moving from flat space to the sphere, because of the conformal nature of these IR and UV fixed points,
we anticipate being able to replace ${\bf p}^2$ and ${\bf p}^{d-2}$ with corresponding GJMS style operators
\cite{GJMS}.  

The general form of the scalar GJMS operator on the unit $d$-sphere is given by (see for example Theorem 2.8 of
\cite{Branson})
\be
\nabla_{2k} = \prod_{j=\frac{d}{2}}^{\frac{d}{2}+k-1} \left( - \Box + j(d-1-j) \right)
\ee
where $\Box$ is the Laplacian on a $d$-dimensional sphere with unit radius.
The eigenvectors are generalized spherical harmonics $Y_{\vec \ell}(\vec \theta)$ labelled by angular quantum numbers $\vec \ell = (\ell_1, \ell_2, \ldots, \ell_d)$ with $|\ell_1| \leq \ell_2 \leq \cdots \leq \ell_d$.  For notational ease, we relabel $\ell_d = \ell$.  
Eigenvectors with the same $\ell$ all have the same eigenvalue.
The eigenvalues of $\Box$ are the standard $-\ell(\ell+d-1)$, which leads to the following result for eigenvalues of $\nabla_{2k}$:
\be
\lambda_\ell = \prod_{j=0}^{k-1} 
\left( \left( \ell + \frac{d-1}{2} \right)^2 - \left(j+\frac{1}{2} \right)^2 \right) \ ,
\ee
with degeneracy
\be
\mbox{deg}(\ell) = \frac{(d+2\ell-1) \Gamma(d+\ell-1)}{\Gamma(d) \Gamma(1+\ell)} \ .
\ee
A generalized scalar field with the equation of motion $\nabla_{2k} \phi = 0$ has scaling dimension $\Delta = \frac{d}{2}-k$.  Written in terms of $\Delta$, we may rewrite the eigenvalues in the form
\be
 \lambda_\ell \equiv  \frac{\Gamma(d+\ell-\Delta)}{\Gamma(\ell + \Delta)} \ .
\ee
 These formulae recover standard results, for example $\lambda_\ell = \ell(\ell+1)$ and $\mbox{deg}(\ell) = 2 \ell+1$ in the case $\Delta =0$ and $d=2$.

We may formally write the log of the partition function for one of these scalar fields in terms of a sum over the eigenvalues:
\be
W_{d, \Delta} = \log Z_{d, \Delta} = -\frac{1}{2} \sum_{\ell=0}^\infty \mbox{deg}(\ell) \log  \left[ \lambda_\ell \left(\frac{R}{\epsilon}\right)^{2\Delta-d}\right] \ .
\ee
We have restored a factor of the size of the sphere $R$ to the eigenvalue, along with a UV regulator $\epsilon$ to keep things dimensionless. More physically, we can think of $W_{d, \Delta}$ as measuring the change in the free energy as the sphere is increased from radius $\epsilon$ to radius $R$.

This sum is strongly divergent in general and requires careful handling.  We begin by examining instead a sum over the derivative with respect to $\Delta$ in the case of interest $d=2$:
\be
\label{mysum}
\partial_\Delta W_{2, \Delta} = \frac{1}{2} \sum_{\ell=0}^\infty(2\ell+1) 
\left( -2 \log \frac{R}{\epsilon} + \psi(\ell +2-\Delta) + \psi(\ell+\Delta) \right) \ ,
\ee
where $\psi(z) = \Gamma'(z)/\Gamma(z)$ is the digamma function.  Truncating at $\ell=\ell_{\rm max}$, the sum can be evaluated, for example using Mathematica \cite{Mathematica}.  We first expand the result at large $\ell_{\rm max}$ and then make the replacement $\ell_{\rm max} = R / \epsilon$.
The result is
\be
\partial_\Delta W_{2, \Delta} = -\frac{R^2}{2 \epsilon^2} - (\Delta-1)^2 \log \frac{R}{\epsilon}  + \ldots \ .
\ee
Integrating this result up yields
\be
W_{2, \Delta}  - W_{2,1} \sim -\frac{(\Delta-1)^3}{3} \log \frac{R}{\epsilon} \ ,
\ee
discarding the regulator dependent $O(\epsilon^{-2})$ power law contribution to the difference.
From this result, it is straightforward to recover (\ref{Wdiff}).
Assuming $W_{2,1} = 0$, we recover the familiar result for a free scalar field $W_{2,0} \sim \frac{1}{3} \log \frac{R}{\epsilon}$
(see for example \cite{Herzog:2015ioa}).  

Other types of regulators give the same result.  Instead of cutting off the sum (\ref{mysum}) at $\ell = \ell_{\rm max}$, one can insert a regulating function 
$\eta(\ell/\ell_m)$ into the sum.  Standard choices are $\eta(x) = x^{-s}$ and $e^{-x}$.  The latter is a member of the family of regulators
\cite{Padilla:2024mkm}, 
 \be
 \eta_{[0,s]}(x) = e^{-x} \sum_{k=0}^{s+1} \binom{s+1}{k} \frac{(-x)^k}{k!}\ ,
 \ee
 inspired by Tao's smoothed asymptotics \cite{tao2013compactness}.  This family gives for any integer $s\ge 1$ the result
 \be
 \partial_\zeta W_\infty  = -\frac{2}{s(s+1)} \ell_m^2 + \frac{1}{s+1} \ell_m  +\zeta^2 \log  \ell_m + \text{finite} \ ,
 \ee
 where we have made the identification $\ell_{m} = R/\epsilon$, giving the same logarithmic dependence as before.

Detailed exposition of this type of difference in sphere partition functions can be found in the AdS/CFT literature, where corresponding renormalization group flows occur due to boundary double trace deformations.  See for example \cite{Gubser:2002zh,Gubser:2002vv,Diaz:2007an,Giombi:2013yva,Giombi:2014xxa}.
There are also deep connections with the boundary conformal field theory literature.  See for example \cite{Herzog:2019bom}.

\bibliographystyle{jhep}
\bibliography{bib.bib}

\end{document}